\documentstyle[prb,aps,epsfig,twocolumn]{revtex}
\begin{document}
\draft
\flushbottom
\twocolumn[
\hsize\textwidth\columnwidth\hsize\csname @twocolumnfalse\endcsname

\title{Quasiclassical Surface of Section Perturbation Theory}
\author{R. E. Prange,\cite{presad} R. Narevich\cite{presad} and Oleg Zaitsev}
\address{Physics Department, University of Maryland, College Park, MD 20742}
\maketitle

\tightenlines
\widetext
\advance\leftskip by 57pt
\advance\rightskip by 57pt
\begin{abstract}
Perturbation theory, the quasiclassical approximation and the quantum
surface of section method are combined for the first time. This solves the
long standing problem of quantizing the resonances and chaotic regions
generically appearing in classical perturbation theory. The result is
achieved by expanding the `phase' of the wavefunction in powers of the {\em 
square root} of the small parameter. It gives explicit WKB-like
wavefunctions even for systems which classically show hard chaos. We also
find analytic solutions to some questions raised recently.
\end{abstract}

\vskip .5cm
\pacs{PACS: 05.45.+b, 03.65.Sq, 72.15.Rn}

]

\narrowtext
\tightenlines
\setcounter{equation}{0}

Perturbation expansions in a small parameter $\epsilon $ are important in
both classical and quantum physics. Not only are valuable approximations
produced, but the breakdown of the expansion can signal new physics.

Poincar\'e found that classical perturbation theory [PT] on an integrable
system fails in two [or more] dimensions for any $\epsilon ,$ due to `small
denominators'. The Kolmogorov-Arnol'd-Moser [KAM]\cite{KAM} theory greatly
illuminated the subject and showed that the breakdown of PT signals chaos.
Phase space trajectories of an integrable system lie on invariant tori.
Under perturbation, periodic orbits, on `rational' tori, are destroyed
except for one or more stable and unstable orbits. The rational tori are
labelled by $p,q$, where $p$ is the winding number and $q$ is the number of
returns to the SS per period. New invariant tori are formed around the
stable orbits while chaos develops near the unstable orbit. The original
tori near the rational one are also destroyed, to a width in action $\sqrt{%
\epsilon }S_{pq}.$ The characteristic action $S_{pq}$ generically drops off
rapidly with $q.$ This is usually pictured, as in Fig.1, on a surface of
section [SS], a slice through the tori, where the structure of alternating
stable and unstable orbits is called an `island chain' or `resonances'.

Quantization of such a system has been of great interest. A rule of thumb is
that only phase space structures of area Planck's constant $h$ or greater
are reflected in the quantum result. Thus, if $\sqrt{\epsilon }S_{pq}<<\hbar 
$ ordinary quantum perturbation theory works well. Above the `Shuryak
border' [SB]\cite{Shuryak}, $\sqrt{\epsilon }S_{pq}\geq \hbar $, ordinary
perturbation theory breaks down as a number $\sqrt{\epsilon }S_{pq}/\hbar $
unperturbed quantum states are strongly mixed by the perturbation. Thus
quantum perturbation theory for small $h$ depends critically on the relation
between $\hbar $, $\sqrt{\epsilon },$ and the torus $p,q.$

Small $h$ is not a perturbation: rather the quasiclassical approximation
[QCA] is used. Combined QCA and PT has been studied over the years: the
perturbed trace formula most recently\cite{Ullmo}. The SB does not appear in 
\cite{Ullmo}, because effectively only short times are considered, where
classical perturbation theory {\em does} work.

We here combine for the first time, PT, the QCA, and the powerful QCA-SS
method of Bogomolny\cite{bogolss,dorsmil}. We achieve quite complete
and explicit results. Namely, we are able to find all the energy levels and
wavefunctions in a WKB approximation for small $\epsilon ,$ provided $h$ is
not too small in a sense to be specified. Interestingly enough, we are led
to express the wavefunctions in terms of a phase $G$ which is generically
expanded as a series in $\sqrt{\epsilon }$ rather than $\epsilon $. This
series is generated in a novel recursive way: A partial $n-1$'th order
solution is obtained which allows a partial $n$'th order solution which in
turn gives the complete $n-1$'th order and a partial $n+1$'th order, and so
on. We also give some numerical checks.

A number of problems susceptible to this new technique have appeared recently%
\cite{Stone,BCL,FS}, and we have also learned of several
ongoing efforts\cite{etal}. These problems are related to a weakly deformed
circular billiard. The Helmholtz equation $\left( \nabla ^2+k^2\right) \Psi
=0$ is to be solved for eigenfunctions $\Psi =\Psi _a$ and eigenvalues $%
k=k_a $ with, say, Dirichlet conditions $\Psi _a(r,\theta )=0$ on the
boundary, $\partial B$. The latter is expressed in polar coordinates by $%
r(\theta )=R_0+\epsilon \Delta R(\theta ).$ These and similar {\em boundary
perturbations} have heretofore been treated \cite{MF} by methods valid only
below the SB.

We illustrate with the stadium billiard\cite{BCL} which has two semicircular
endcaps of radius $R_B$ connected by parallel straight sides of length $2a$.
Then $\epsilon =a/R_B$ is assumed small and $R_0$ is taken as $R_0\approx
R_B+2a/\pi $ while $\Delta R/R_0\approx \left| \sin \theta \right| -2/\pi .$
This `stadium' choice of $\Delta R$ has a discontinuous first derivative so
KAM does not apply. Fig. 1a shows that the classical map deviates from
the new invariant tori [given approximately by $l_{WKB} (\theta)$
defined below] after about $1/\sqrt{\epsilon}$ iterations. We also
show results, Fig. 1b, for a `smoothed stadium',
a truncated Fourier series of the `stadium' $\Delta R(\theta )$, where
the orbit stays on a new invariant torus.

Classically the stadium is chaotic with {\em no} stable orbits. Orbits {\em %
diffuse} in angular momentum at{\em \ long times}\cite{BCL}. It was thought
that such hard chaos systems do not have simple, analytically expressible
wavefunctions when quantized. Thus, qualitative and statistical questions,
such as the existence and statistics of localization, have been considered.
Our explicit analytic results were therefore quite unexpected, and we are
able to interpret the results directly in terms of analytic wave functions.
The results are possible because it is the {\em short time, nearly regular}
behavior which determines the quantization.

In quantum language, we take units $R_0=1,$ $\hbar =1,$ particle mass = $1/2$%
, so $k$ is the dimensionless wavenumber, [equivalent to $1/\hbar$]. 
We take the
billiard boundary $\partial B$ as SS. Then Bogomolny's unitary operator is%
\cite{bogolss} 
\begin{equation}
T(\theta ,\theta ^{\prime },k)=-\left( \frac k{2\pi i}\frac{\partial
^2L(\theta ,\theta ^{\prime })}{\partial \theta \partial \theta ^{\prime }}%
\right) ^{\frac 12}\exp \left( ikL\left( \theta ,\theta ^{\prime }\right)
\right)  \label{T}
\end{equation}
where $L$ is the chord distance between two points, specified by polar
angles, on $\partial B$. Expanding, 
\begin{eqnarray}
kL(\theta ,\theta ^{\prime }) &=&2k\left| \sin \frac{\theta -\theta ^{\prime
}}2\right| \left( 1+\epsilon \frac{\Delta R(\theta )+\Delta R(\theta
^{\prime })}2\right) +\ldots  \nonumber  \label{LT} \\
\ &=&k(L_0+\epsilon L_2+\ldots .)  \label{LT}
\end{eqnarray}

The energy levels [$k_a$] of the system are given in QCA\cite{bogolss} by
solution of $\det (1-T(k))=0$. Our seemingly more difficult technique
studies 
\begin{equation}
\psi (\theta )=\int d\theta ^{\prime }T(\theta ,\theta ^{\prime },k)\psi
(\theta ^{\prime }),  \label{psi}
\end{equation}
solvable only for $k=k_a.$ [$\psi \approx \partial \Psi /\partial n$ on $%
\partial B$.]

We start with the Ansatz $\psi (\theta )=\exp \left( i\alpha f(\theta
)\right) $ where $df/d\theta =$ $f^{\prime }$ $\sim 1$ and $k>>\alpha
>>k\epsilon $. This Ansatz represents a superposition of angular momentum
states $\left| l\right| \sim \alpha <<l_{\max }=k,$ and for $\alpha >1$
conveniently expresses the mix of states needed to diagonalize the
Hamiltonian above the SB.

\begin{figure}[tbp]
{\hspace*{-.5cm}\psfig{figure=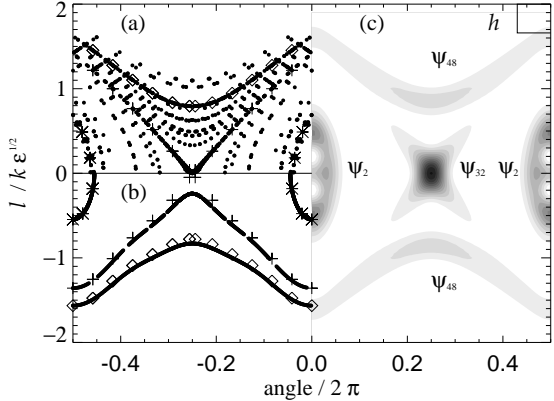,height=6cm}} \vspace*{.12in}
\caption{SS [angular momentum vs angle] of orbits for a) `stadium' and b)
`smoothed stadium' $\epsilon =0.01$. Points on $l_{WKB}(\theta ),\,$ for
`continuum',$\Diamond $, `separatrix',$+,$ and `bound', $*$, values of $E_m,$
respectively. Three orbits, each iterated 1000 times, coalesce into solid
lines in b), where KAM applies. Orbits started at the symbols iterated
forward and backward 15 times appear as dots in a) where KAM fails. Only
short time structure is regular. c) Husimi plots for exact states of
Fig. 2, as well as a `scar' state $\psi _{32}$. Square has area $h.$} \label
{fig:1}\end{figure}

According to the stationary phase [S$\Phi $] method, the $\theta ^{\prime }$
integral is dominated by the region $\theta ^{\prime }\sim \theta \pm \pi $
where $kL_0=2k\left| \sin \frac 12(\theta -\theta ^{\prime })\right| $ is
stationary.$\,$ Expand $\sin \frac 12(\theta -\theta ^{\prime })\approx 1-%
\frac 18\delta \theta ^2$ , $\delta \theta =\theta ^{\prime }-\theta -\pi $
to find 
\begin{equation}
S(\theta ,\theta ^{\prime })=kL\approx 2k-%
{\textstyle {1 \over 4}}
k\delta \theta ^2+k\epsilon (\Delta R(\theta )+\Delta R(\theta ^{\prime }))
\label{kLa}
\end{equation}
[In Eq.(\ref{kLa}) we replaced $L_0$ by $2$, its stationary value, when
multiplied by $\epsilon .$]

Regarding $S$ as a classical generating function, we obtain the surface of
section maps $(l^{\prime },\theta ^{\prime })\rightarrow (l,\theta )$ found
earlier\cite{BCL},\cite{FS} by $l=\partial S/\partial \theta ,$ $l^{\prime
}=-\partial S/\partial \theta ^{\prime }$. Motivated by this, Borgonovi\cite
{etal} has studied the $T$ operator and classical map given by Eq.(\ref{kLa}%
) with $\Delta R=\left| \sin \theta \right| $ and $\delta \theta =\theta
^{\prime }-\theta $ . This system is `almost' the quantum kicked
rotor-classical standard map\cite{Chirikov}, which corresponds to $\Delta
R=\sin \theta $. Thus, in addition to solving the distorted billiard
problem, we can also solve an important class of quantized perturbed twist
maps.

Returning to Eq.(\ref{psi}), we expand all functions of $\theta ^{\prime }$
about $\theta +\pi .$ I.e. $f(\theta ^{\prime })\approx f(\theta +\pi
)+\delta \theta \,f^{\prime }(\theta +\pi )$ to order $\delta \theta $,
[since $\alpha <<k$] and $\Delta R(\theta ^{\prime })\approx \Delta R(\theta
+\pi ),$ since $\alpha >>k\epsilon $. Doing the integral reduces Eq.(\ref
{psi}) to 
\begin{eqnarray}
&&\exp [i\alpha f(\theta )]  \nonumber  \label{E1} \\
&=&i\exp \left[ i\left( 2k+(\alpha f^{\prime })^2/k+k\epsilon V(\theta
)+\alpha f(\theta +\pi )\right) \right]  \label{E1}
\end{eqnarray}
where $V(\theta )=\Delta R(\theta )+\Delta R(\theta +\pi ).$

For Eq.(\ref{E1}) to hold, the exponents of order $\alpha $ must combine to
give a constant $c$, i.e. $f(\theta +\pi )=f\left( \theta \right) +c.$ Now
take $\alpha =k\sqrt{\epsilon }$ so a solution is possible provided $%
(f^{\prime })^2+V(\theta )$ is a constant, which we call $E_m$. Thus 
\begin{equation}
f(\theta )=\pm \int_0^\theta d\theta ^{\prime }\sqrt{E_m-V(\theta ^{\prime })%
}  \label{f}
\end{equation}
reminiscent of elementary WKB theory. The constant of integration is
irrelevant. Notice $V(\theta )=V(\theta +\pi )\Longrightarrow f(\theta +\pi
)=f\left( \theta \right) +c.$ We define $l_{WKB}(\theta )=k\sqrt{\epsilon }%
f^{\prime }(\theta ).$

Assuming for now that $E_m\geq V$ , [a `continuum' state], we must choose $%
E_m$ such that $kb\int_0^{2\pi }d\theta ^{\prime }\sqrt{E_m-V(\theta
^{\prime })}=2\pi m$ where $m$ is integer, $b=\sqrt{\epsilon }$ and so $%
c=\pi m/kb.$ The condition giving the energy is 
\begin{equation}
\exp \left[ i\left( 2k+kb^2E_m+kbc+\pi /2\right) \right] =1=\exp (2\pi in)
\label{eigenv}
\end{equation}
which has solutions $k=k_{n,m}.$ For $\Delta R=0$, this reduces to $%
2k+m^2/k+\pi m=(n-\frac 14)2\pi $ equivalent to Debye's approximation to
Bessel's function, valid for $k$ large and $m/k$ small. Thus, this Ansatz
produces states labelled $m,n$ with $m$ an integer angular quantum number
satisfying $\left| m\right| <<k_{n,m}\approx \pi n.$ There are three
symmetries, reflections about the two principal axes and time reversal,
which allows real wavefunctions. Thus the even-even states are $\psi =\cos
\alpha f(\theta )$ and $m$ is an even integer. [We choose the lower limit in
Eq.(\ref{f}) to be at a minimum of $V$.] This result allows an explicit
estimate of $\psi _l$ [angular momentum representation] which, for $\left|
l\right| >k\sqrt{\epsilon }$, decays exponentially for smooth $V$ and as $%
l^{-4}$ for the stadium case. This localization was first\cite{BCL} thought
to be dynamical localization analogous to Anderson localization\cite{FGP},
but now\cite{etal} [for $k\epsilon ^2<1$] is attributed to Cantori\cite
{Cantori}.

If $E_m-V$ changes sign there are `bound state' regions near the minima of $%
V $ [at $\theta =0$] where $E_m>V$, {\em e.g. } let the region be $\left|
\theta \right| <$ $\theta _m<\pi /2$, where $\theta =\theta _m$ is a
`classical turning point' of the motion. The even-even quantization
condition is now, approximately, $\cos (\alpha f(\theta _m))=0,$ or $\alpha
f(\theta _m)=(m+\frac 12)\pi $ and $\psi \approx 0$, $\theta _m<\left|
\theta \right| <$ $\pi /2.$ In this approximation there is a degeneracy
between even-even and even-odd symmetry. This treatment neglects tunnelling
into the forbidden region $V>E_m,$ as well as effects on the amplitude of
the wavefunctions.

The bound states quantize the stable resonance islands and the continuum
states quantize the unstable and perturbed KAM regions. A minimum in $V$ is
at a stable periodic orbit, and a maximum at an unstable one. More
correctly, if $V$ does not have sufficiently many derivatives, the stable
orbits can disappear, but the quantum system is hardly affected, if the
wavelength is not too short. `Scars' of unstable orbits, Fig. 1c, are states
with $E_m$ just greater than the maximum $V.$ Fig. 2 shows a WKB state and
two indistinguishable numerically obtained states, all with the same value
of $k\sqrt{\epsilon },$ one with $k\epsilon =1.8$ the other for $k\epsilon
=0.18.$ This shows the state depends dominantly on $k\sqrt{\epsilon }$ and
suggests no qualitative changes occur at $k\epsilon \approx 1$. Husimi plots
of these states are shown in Fig. 1c.

Borgonovi\cite{etal} has numerically calculated an average localization
width in angular momentum, $l_\sigma $, where, [in effect] $l_\sigma
^2=\sum_a|c_a|^2\int d\theta \left| \psi _a^{\prime }\right| ^2$ and $c_a$
is the normalized zero'th Fourier component of the eigenstate $\psi _a.$
According to the results just obtained, the $c_a$'s should be relatively
small for `continuum' states, [Fig. 2] since the phase is not stationary, so
the `bound' states dominate. Then $l_\sigma ^2\approx \alpha
^2\sum_a|c_a|^2\int d\theta \left( E_{m(a)}-V(\theta )\right) $ $\left| \sin
\alpha f_a(\theta )\right| ^2$ The sum is now a sort of average $\left(
E-V\right) $ which is of order unity and nearly independent of $\alpha .$
Thus $l_\sigma \propto k\sqrt{\epsilon }$. Borgonovi fixed $k$ and increased 
$\epsilon $, agreeing with this result until $k\epsilon ^2\approx 1.$ We
show below that our theory should fail at that point. We note that the
result depends on the definition of the $c_a$'s. The result can be quite
different if the $c_a$'s are chosen to be the overlap of $\psi _a$ with some
high angular momentum state, for example. Fig. 2 shows a high angular
momentum state, away from a resonant torus, which has a much narrower
distribution. [See the next paragraph.]

\begin{figure}
{\hspace*{-0.2cm}\psfig{figure=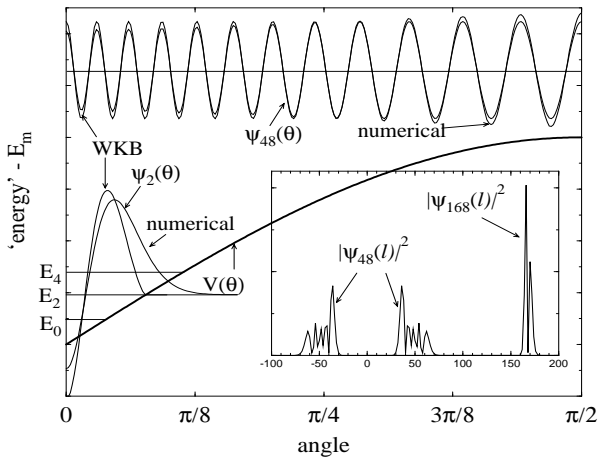,height=8.cm,width=6cm,angle=270}} 
{\vspace*{.31in}}
\caption{`Stadium' potential $\left| \sin \theta \right| $ vs angle. States
and potential are symmetric about zero angle. Bound and continuum, WKB and
exact states are shown, with zeroes at WKB `energy' parameter $E_m.$ 
$k$${\epsilon }^{1/2}$ $=42.3$ 
is fixed.  
Inset: Angular momentum representation of
continuum state $m=48$ and  exact state near angular momentum $m=168.$}
\label{fig:2}
\end{figure} 

We turn to general angular momenta and higher orders in $\epsilon .$ We look
for solutions of the form $\psi =\exp (iG(\theta ^{\prime }))$, where $%
G=l\theta ^{\prime }+k(\epsilon f_2+\epsilon ^2f_4+\ldots ).$ The $f$'s are $%
2\pi $-periodic and $l\leq k$ is integer. This, if successful, is a usual PT
for $G$. The $S\Phi $ angle is $\theta ^{\prime }\approx \theta +\Theta _l$
where $\Theta _l=-2$sign$(l)\cos ^{-1}(l/k).$ Expanding as before the order $%
k\epsilon $ condition is 
\begin{equation}
f_2(\theta +\Theta _l)-f_2(\theta )=\tilde L_2(\theta ).  \label{df2}
\end{equation}
We use $\tilde l_2$ and $\tilde L_2(\theta )$ as the constant and variable
parts of $L_2(\theta ,\theta +\Theta _l).$ The constant part $k\epsilon 
\tilde l_2$ contributes to the phase of Eq.(\ref{eigenv}). Eq.(\ref{df2}) is
solved in terms of $r\neq 0$ Fourier components, i.e. $f_{2r}=\left( \exp
(ir\Theta _l)-1\right) ^{-1}\tilde L_{2r}.$ This a good solution unless the
denominator is excessively small. It never strictly vanishes since $\Theta
_l/2\pi $ is an irrational number. However, if $\Theta _l$ is close to $%
\Theta _{pq}=2\pi p/q$, where $p/q$ is a rational number, [corresponding to
the strongly perturbed rational tori of classical perturbation theory], then
the denominator will be small if $r$ is a multiple of $q.$ It will still be
a good solution if $\tilde L_{2r\text{ }}$vanishes or is sufficiently small.
Generically $\tilde L_{2r}$ decreases rapidly for large $r.$ If the small
denominators are thus compensated by small numerators, this perturbation
theory can be carried to higher orders by the methods described below. If
not, we need to refine the approach along the lines of our first Ansatz
which corresponds to $q=2.$ This small denominator problem is the analog in
QCA of the small denominator problem of classical PT\cite{KAM}.

We are thus motivated to consider 
\begin{equation}
\psi =\exp \left[ i\left( l_{pq}\theta ^{\prime
}+k(bf_1+b^2f_2+b^3f_3+\ldots )\right) \right]  \label{A2}
\end{equation}
The [non-integer] angular momentum $l_{pq\text{ }}\,$is chosen to make the
stationary point $\theta ^{\prime }=\theta +\Theta _{pq}.$ Expanding as
before, the order $b$ requirement is $f_1(\theta +\Theta _{pq})-f_1(\theta
)+l_{pq}\Theta _{pq}/kb=c=$constant implying $f_1^{\prime }$ is $q$%
-periodic, i.e. periodic with period $\Theta _{pq}.$ At order $b^2$ we have 
\begin{equation}
S_{pq}^{-1}\left( f_1^{\prime }\right) ^2+\tilde L_2(\theta )+f_2(\theta
+\Theta _{pq})-f_2(\theta )=E_m  \label{delf2}
\end{equation}
where $\tilde L_2(\theta )$ is the variable part of $L_2(\theta ,\theta
+\Theta _{pq}),$ $S_{pq}=\left| \sin \frac 12\Theta _{pq}\right| $ and $E_m$
is to be determined. We divide Eq.(\ref{delf2}) into $q$-periodic and non $q$%
-periodic parts. The nonperiodic terms $f_{2\text{ }}$and $\tilde L_2$ must
combine to give a $q$-periodic result, thus 
\begin{equation}
f_2(\theta +\Theta _{pq})-f_2(\theta )+\tilde L_2(\theta )=\bar V_q(\theta )
\label{delf2b}
\end{equation}
where $\bar V_q(\theta )$ is to be determined. We `$q$-average' both sides
giving $\bar V_q(\theta )=\frac 1q\sum_{j=1}^q\tilde L_2(\theta +j\Theta
_{pq}).$ Expressed in Fourier components, $\bar V_q(\theta )=\sum_l\tilde L%
_{2ql}e^{iql\theta }\,$ and $f_2(\theta )=\sum_l^{\prime }(1-e^{il\Theta
_{pq}})^{-1}\tilde L_{2l}e^{il\theta }+\bar f_2(\theta )$. The prime
indicates that integers $l$ divisible by $q$ are not included in the sum and 
$\bar f_2(\theta )$ is an $q$-periodic function not yet determined. Then 
\begin{equation}
f_1(\theta )=\pm S_{pq}^{1/2}\int_0^\theta d\theta ^{\prime }\sqrt{E_m-\bar V%
_q(\theta ^{\prime })}  \label{f1r}
\end{equation}
and considerations like those discussed earlier for $q=2$ fix the
quantization of $E_m.$ The size of $\bar V_q$, which decreases rapidly with $%
q,$ determines if powers of $\sqrt{\epsilon \text{ }}$ rather than $\epsilon 
$ are needed.

Order $b^3$ is more complicated: $L_2(\theta ,\theta ^{\prime })$ and $f_2$
are expanded to $\delta \theta $, $f_1$ to $\delta \theta ^2$ and $L_0$ to $%
\delta \theta ^3$. The integral of Eq.(\ref{psi}) is thus 
\begin{equation}
\int d\delta \theta \exp \left[ \frac{-ik}{3!}L_0^{\prime \prime \prime
}\delta \theta ^3+\frac{ik}2(L_0^{\prime \prime }+bf_1^{\prime \prime
})\delta \theta ^2+iF^{\prime }\delta \theta \right]  \label{int3}
\end{equation}
where $F^{\prime }=kbf_1^{\prime }+kb^2F_2^{\prime }$ with $F_2^{\prime
}=f_2^{\prime }+L_2^{\prime }.$ We denote derivatives evaluated at $\theta
^{\prime }=\theta +\Theta _{pq}$ by primes. [This integral is done over a
region near the original stationary point. The new stationary point coming
from $\delta \theta ^3$ is not meaningful.] The width of contributing angles 
$\delta \theta $ is of order $k^{-1/2}$ which is small. However, the shift
of the center of the contributing region is expressed by a power series in $%
b $ whose leading term is $-bf_1^{\prime }/L_0^{\prime \prime }$. If $%
kb^2\geq 1$, the shift cannot be neglected. Thus, to order $b^3$ we require 
\begin{eqnarray}
&&\frac{L_0^{\prime \prime \prime }}{3!}\left( \frac{-f_1^{\prime }}{%
L_0^{\prime \prime }}\right) ^3-\frac 12f_1^{\prime \prime }\left( \frac{%
f_1^{\prime }}{L_0^{\prime \prime }}\right) ^2-\frac{f_1^{\prime
}F_2^{\prime }}{L_0^{\prime \prime }}+c_3  \nonumber  \label{delf3} \\
&=&-f_3(\theta +\Theta _{pq})+f_3(\theta )  \label{delf3}
\end{eqnarray}
where $c_{3\text{ }}$is a constant. Let $F_2^{\prime }=\bar f_2^{\prime
}+A(\theta ),$ where $A$ has already been determined by lower order
considerations. Eq.(\ref{delf3}) can only be satisfied if the $q$-average of
the left hand side vanishes. This determines $\bar f_2^{\prime }$ by 
\[
\bar f_2^{\prime }=-\bar A_q+\frac{L_0^{\prime \prime }c_3}{f_1^{\prime }}-%
\frac 12f_1^{\prime \prime }\frac{f_1^{\prime }}{L_0^{\prime \prime }}+\frac{%
L_0^{\prime \prime \prime }}{3!}\left( \frac{f_1^{\prime }}{L_0^{\prime
\prime }}\right) ^2. 
\]
This expression must also have vanishing angular average, since $\bar f%
_2^{\prime }$ is the derivative of a periodic function, which determines $%
c_3.$ Thus $f_2$ is determined up to an irrelevant integration constant,
and, then as before, $f_3$ is determined up to an $q$-periodic function.

If $kb^3<<1,$ we may stop here. If not, we can continue finding higher order
corrections, expanding to higher powers of $\delta \theta $ and keeping the
terms $L_4$, $L_{6,}\,\ldots $ in the expansion of the phase of the $T$
operator. The series will be effectively terminated at order $n$ when $%
kb^n<<1.$ However, the method may break down sooner, indicating a change in
the fundamental physics.

For example, `bound state' solutions of Eq.(\ref{f1r}) give infinite second
derivatives $f_1^{\prime \prime }$ when the square root vanishes, i.e. at
the `classical turning points'. This, however, can be taken into account to
give the familiar turning point corrections of elementary WKB theory.

In the case $\Delta R=\left| \sin \theta \right| $, there are $\delta $%
-function singularities in $f_1^{\prime \prime \prime }$ and $L_2^{\prime
\prime }$. These large derivatives invalidate the expansion. Thus in the
Bunimovich problem we expect our solution to break down when $k\epsilon
^2>1. $ Numerical results\cite{etal} seem to confirm this expectation,
giving two different behaviors in either side of this border.

In principle, we can use this technique to study perturbations of any two
dimensional integrable system. `Simply' use action angle coordinates $I_1,$ $%
I_2$, $\Theta _1$, $\Theta _2$, and take as surface of section $\Theta _1=0$%
. The $T$ operator will have a phase $k(S_0(\Theta _2-\Theta _2^{\prime
})+\epsilon S_2(\Theta _2,\Theta _2^{\prime })+\ldots )$ and the rest is
pretty much the same as above. Other coordinates may be more convenient in
practice, however. The circle is nice because the action-angle coordinates
are immediate.

There are other applications of this technique in nonperturbative settings,
in which certain classes of eigenstates can be found. The germ of the method
first appeared in the study of the ray splitting billiard\cite{ABGOP}, and
it can be used to find the well known `bouncing ball' states in the [large $%
\epsilon $] stadium billiard.

We have thus produced a fairly general theory allowing us to find the effect
of perturbations on integrable quantum systems which exploits the
quasiclassical approximation and the surface of section technique. If the
perturbation classically gives rise to resonances big enough to influence
the quantum problem, we must expand in the square root of the small
parameter. If the resonances are small, a simpler expansion works.

Supported in part by NSF DMR 9624559 and the U.S.-Israel BSF 95-00067-2. We
thank the Newton Institute for support and hospitality. Many valuable
discussions with the organizers and participants of the Workshop `Quantum
Chaos and Mesoscopic Systems' contributed to this work.

\end{document}